\documentclass[aps,pre,twocolumn,superscriptaddress,groupedaddress]{revtex4-1}

\usepackage{graphicx}
\usepackage{mathrsfs} 

\usepackage{dcolumn}   
\usepackage{bm}        
\usepackage{amssymb}
\usepackage{amsmath}
\usepackage{lipsum}
\usepackage{enumitem} 
\usepackage{color}

\usepackage[normalem]{ulem}

\begin{document}


\title{Persistence of Non-Markovian Gaussian Stationary Processes in Discrete Time}

\author{Markus~Nyberg} \email{markus.nyberg@umu.se} \affiliation{Integrated Science Lab, Department of Physics, Ume\r{a} University, SE-901 87 Ume\r{a}, Sweden}
\author{Ludvig~Lizana} \affiliation{Integrated Science Lab, Department of Physics, Ume\r{a} University, SE-901 87 Ume\r{a}, Sweden} 
 
\date{\today}


\begin{abstract}
The persistence of a stochastic variable is the probability that it does not cross a given level during a fixed time interval. Although persistence is a simple concept to understand, it is in general hard to calculate. Here we consider zero mean Gaussian stationary processes in discrete time $n$. Few results are known for the persistence $P_0(n)$ in discrete time, except the large time behavior which is characterized by the nontrivial constant $\theta$ through $P_0(n)\sim \theta^n$. Using a modified version of the Independent Interval Approximation (IIA) that we developed before, we are able to calculate $P_0(n)$ analytically in $z$-transform space in terms of the autocorrelation function $A(n)$. If $A(n)\to0$ as $n\to\infty$, we extract $\theta$ numerically, while if $A(n)=0$, for finite $n>N$, we find $\theta$ exactly (within the IIA). We apply our results to three special cases: the nearest neighbor-correlated "first order moving average process" where $A(n)=0$ for $ n>1$, the double exponential-correlated "second order autoregressive process" where $A(n)=c_1\lambda_1^n+c_2\lambda_2^n$, and power law-correlated variables where $A(n)\sim n^{-\mu}$. Apart from the power-law case when $\mu<5$, we find excellent agreement with simulations.
\end{abstract}

\maketitle

\textbf{\emph{Introduction}.---}In this rapid communication we study a Gaussian stationary process (GSP) $x(n)$ of zero mean in discrete time $n$. We are interested in the persistence probability $P_0(n)$, which is the probability that $x(n)$ has not changed sign up to step $n$. 

Even though simple to understand, it is in general challenging to calculate $P_0(n)$ exactly. Even after decades of efforts by  mathematicians \cite{rice1958distribution,mandelbrot1968fractional,blake1973level,slepian1962one,burkhardt1993semiflexible,sinai1992distribution} and theoretical physicists \cite{derrida1994non,derrida1995exact,majumdar1996survival,majumdar1996nontrivial,krapivsky1996life,krug1997persistence,majumdar1996global,derrida1996persistent,kallabis1999persistence,krishnamurthy2003persistence}, the problem remains unsolved. Historically the field was theoretically driven, but more recently  several experimental groups also contributed with new insights \cite{marcos1995self,tam1997first,tam2002cluster,yurke1997experimental,wong2001measurement,soriano2009universality,takeuchi2012evidence}. For example from  measuring  the decay time of clusters in soap froth \cite{tam2002cluster}, and the mean spin magnetization in a laser-polarized Xenon gas \cite{wong2001measurement}.

On the theoretical side, most results come from studies of continuous time processes (see \cite{bray2013persistence} for a comprehensive review). However,  these results do not simply generalize to discrete time processes, which means that they  cannot  be applied to time series data coming from measurements or simulations. In this paper we narrow this gap.  Specifically for GSPs.

To derive our results, we used the Independent Interval Approximation (IIA) \cite{mcfadden1958axis,sire2008crossing}, that we recently generalized to handle GSPs in discrete time \cite{nyberg2016simple,nyberg2017}. In short, the IIA splits the total observation time into intervals, where the endpoints of the intervals correspond to sign changes of $x(n)$. Then we assume that the lengths of these intervals are uncorrelated with each other. This converts $x(n)$ into a 'clipped' process, where the memory is erased at every sign change. Indeed, this is an inaccurate treatment if the processes' memory extends over several intervals, for example for power law correlated variables, but as we demonstrate, it works well for processes with finite memory.

Based on our method, we derive $P_0(n)$ analytically in $z$-transform space (a discrete Laplace transform), as well as a recursion relation in time domain. To evaluate our expressions, we only need to specify the process' autocorrelation function.

Furthermore, for the simplest GSPs, the Markovian Ornstein-Uhlenbeck process \cite{majumdar2001persistence} and the non-Markovian random acceleration process  \cite{ehrhardt2002persistence}, we know that $P_0(n) \sim\theta^n$ for large times $n$. Here $\theta$ is the persistence constant, which depends non-trivially on the autocorrelator. To find $\theta$ for any GSP,  we derive a semi-analytic expression in terms of the autocorrelation function. We show that our formula works well when the correlation between variables decays exponentially or when it is nearest neighbor-correlated. In summary,  we find simple yet accurate results for: 
\begin{enumerate}
\item The persistence probability $P_0(n)$ for $n\geq0$ through a recursive relation.
\item The persistence constant $\theta$ via a summation formula, that can be solved analytically for nearest neighbor-correlated variables.
\end{enumerate}
As a sub result, we also calculate the mean first-passage time till the first sign change using a summation formula.

\begin{figure}[] 
	\centering
	\includegraphics[width=1\columnwidth]{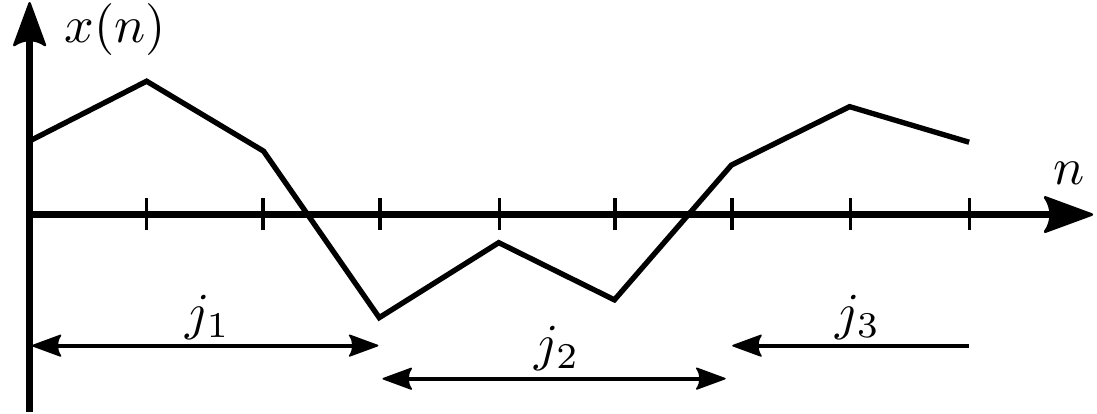}
	\caption{Stochastic time series describing the continuous position $x(n)$ as a function of the discrete time $n$. Time intervals $T_1,T_2,\ldots$ denote times spent above and below the origin. Note that after a sign change, the position is not necessarily zero. Hence, the subsequent sign change can be viewed as a first passage to zero starting from this position. \label{fig:x(n)}}
\end{figure}

\textbf{\emph{Derivation of equations}.---}To calculate the persistence $P_0(n)$, we first find the first-passage time density (FPTD), $\rho(n)$, using the IIA. The FPTD is related to the persistence via $P_0(n)=1-\sum_{k=0}^n\rho(k)$, which can be re-written as
\begin{equation} \label{eq:persist+fptd} 
P_0(n)=P_0(n-1)-\rho(n).
\end{equation}
The persistence is a special case of the more general probability that $m$ sign changes occur up to $n$. Denoting this by $P_m(n)$, we start by splitting the total observation time $n$ into $m$ intervals, see Fig. \ref{fig:x(n)}. The first time interval $j_1$ is the first-passage time and thus, related to the FPTD $\rho(j_1)$. The subsequent intervals are drawn from the first-return density $\psi(j)$, where we assume that the process $x(n)$ has the same dynamics on both sides of the origin. Indeed, this is an approximation for processes with memory. But for a zero mean GSP, the probability of being above or below the origin is 1/2, which tells us that there should not be any significant difference between any two consecutive intervals $j_{i}$ and $j_{i+1}$ for $i>1$. Using the IIA, we can formally write $P_m(n)$ as \cite{nyberg2017} 
\begin{equation} \label{eq:pmn+iia}
\begin{aligned}
P_{m}(n) &= \sum_{j_1=0}^{n}\rho(j_1)\sum_{j_2=j_1}^{n}\psi(j_2-j_1)\sum_{j_3=j_2}^{n}\psi(j_3-j_2)\cdots \\
&\cdots \sum_{j_{m}=j_{m-1}}^{n}\psi(j_{m}-j_{m-1})Q(n-j_{m}),
\end{aligned}
\end{equation}
valid for $m>0$, where $Q(n)=\sum_{j=n+1}^{\infty}\psi(j)$ makes sure that no further sign change occurs after the $m$th crossing. 

The name 'first-return density' is somewhat misleading, since $\psi(n)$ describes the first passage to zero from some position close to zero (see Fig. \ref{fig:x(n)}). However, as shown before \cite{nyberg2016simple,nyberg2017}, this only weakly affects the results coming out of the IIA.

To proceed, we work in $z$-transformed space. The $z$-transform of $f(n)$ is $f(z)=\sum_{n=0}^{\infty}f(n)z^{-n}$ \cite{debnath2014integral}. Applying this to Eq. \eqref{eq:pmn+iia} gives
\begin{equation} \label{eq:pmn2}
P_{m}(z) = \rho(z)\psi(z)^{m-1}\frac{z\left[ 1-\psi(z)  \right]}{z-1}.
\end{equation}

To reduce the number of unknowns, we first use Rice's formula, that gives the mean number of sign changes up to time $n$ for a GSP in discrete time, $\langle m(n) \rangle=nr$ \cite{rice1944mathematical} where $r\equiv \cos^{-1}(A(1))/\pi$ is the rate of sign changes and $A(n)$ is the autocorrelator $\langle x(n+j)x(j) \rangle/\langle x(0)^2 \rangle$. In $z$-transformed space, the Rice formula is given by
\begin{equation} \label{eq:rice}
\langle m(z) \rangle= \frac{zr}{(z-1)^2}.
\end{equation}
Calculating the first moment from Eq. \eqref{eq:pmn2} and using Eq. \eqref{eq:rice} yields the relation
\begin{equation} \label{eq:rel1}
\rho(z)=r\frac{1-\psi(z)}{z-1}.
\end{equation}
Next, if the probability for an odd number of sign changes up to time $n$ is given by $\omega(n)$, then
\begin{equation} \label{eq:pmn6}
\omega(z)=\sum_{m=1}^{\infty}P_{2m-1}(z) = \frac{z\rho(z)}{z-1}\frac{1}{1+\psi(z)},
\end{equation}
where we used Eq. \eqref{eq:pmn2} and summed the geometric series. To solve for the FPTD, we use Eqs. \eqref{eq:rel1} and \eqref{eq:pmn6}. This gives
\begin{equation}\label{eq:fptd+z}
\rho(z)= \frac{2r(z-1)\omega(z)}{rz+(z-1)^2\omega(z)}.
\end{equation}

A formal solution to the FPTD in $n$-space is given by the simple recursive formula (see Appendix \ref{app:a} for details)
\begin{align} \label{eq:fptd+n}
\rho(n+1) &= \Delta\omega(n) - \frac{1}  {2r}\sum_{j=0}^{n}\rho(j)\Delta^2\omega(n-j),
\end{align}
where $\Delta\omega(n)=\omega(n+1)-\omega(n)$. Using Eqs. \eqref{eq:persist+fptd} and \eqref{eq:fptd+n} gives the persistence
\begin{align} \label{eq:persist+n}
P_0(n+1) &=P_0(n)- \Delta\omega(n) - \frac{1}  {2r}\sum_{j=0}^{n}\Delta P_0(j)\Delta^2\omega(n-j),
\end{align}
which only depends on the autocorrelator $A(n)$ through (see Appendix \ref{app:c} for details)
\begin{equation} \label{eq:omega_n}
\omega(n)=\frac{1}{2}-\frac{1}{\pi}\sin^{-1}(A(n)),
\end{equation}
where we note that $\omega(1)=r$.

In summary, Eqs. \eqref{eq:fptd+n} and \eqref{eq:persist+n} are exact  within the IIA and simple to evaluate numerically. They are valid for all $n\geq1$ with the initial conditions $P_0(n)=1$ and $\rho(n)=0$ for $n\leq0$.  The only input is the autocorrelator $A(n)$ that enters in $\omega(n)$.  However, Eq. \eqref{eq:persist+n} is not on the best analytical form to find the persistence constant $\theta$, which characterizes the long-time behavior of $P_0(n)$. Therefore, one must  work in a different direction.

\textbf{\emph{Persistence constant}.---}For large $n$, we assume that the persistence of a zero mean GSP in discrete time decays exponentially for large times $n$ as $P_0(n) \sim \theta^n$ \cite{majumdar2001persistence}, which in $z$-space reads \cite{note1}
\begin{align} \label{eq:persist+n3}
P_0(z)\sim\frac{z}{z-\theta},
\end{align}
where we note that $\theta$ is a pole in $P_0(z)$. Using Eqs. \eqref{eq:persist+fptd} and \eqref{eq:fptd+z}, the persistence in $z$-sapce becomes
\begin{equation} \label{eq:persist+z}
P_0(z)=\frac{z}{z-1}\cdot\frac{rz+(z-1)(z-1-2r)\omega(z)}{rz+(z-1)^2\omega(z)}.
\end{equation}
We then find $\theta$ by solving for the largest root ($<1$) in the denominator of Eq. \eqref{eq:persist+z}, that is,
\begin{equation} \label{eq:roots+theta}
rz^*+(z^*-1)^2\omega(z^*)\big|_{z^*=\theta}=0,
\end{equation}
with the $z$-transform of $\omega(n)$ is given by
\begin{equation} \label{eq:om2+z}
\omega(z)=\frac{z}{2(z-1)}-\frac{1}{\pi}\sum_{n=0}^{\infty}\sin^{-1}(A(n))z^{-n}.
\end{equation}
In general, this sum can not be carried out analytically. 
Therefore, one must truncate the sum at some large value of $n$ when it has converged, and solve Eq. \eqref{eq:roots+theta} numerically. However, there are special cases that we will consider where the sum can be computed analytically that yields $\omega(z)$ on closed form and therefore an analytical expression for $P_0(n)$ and $\theta$. 

\textbf{\emph{Simulations and results}.---}With the theory laid out, we now turn to applications. We will consider three different non-Markovian GSPs and compare them to simulations and literature results when possible. When simulating the GSP, we use the algorithm in \cite{wood1994simulation} that generates random trajectories based on the two-point correlator $\langle x(n+j)x(j) \rangle$.

\begin{figure}[] 
	\centering
	\includegraphics[width=1\columnwidth]{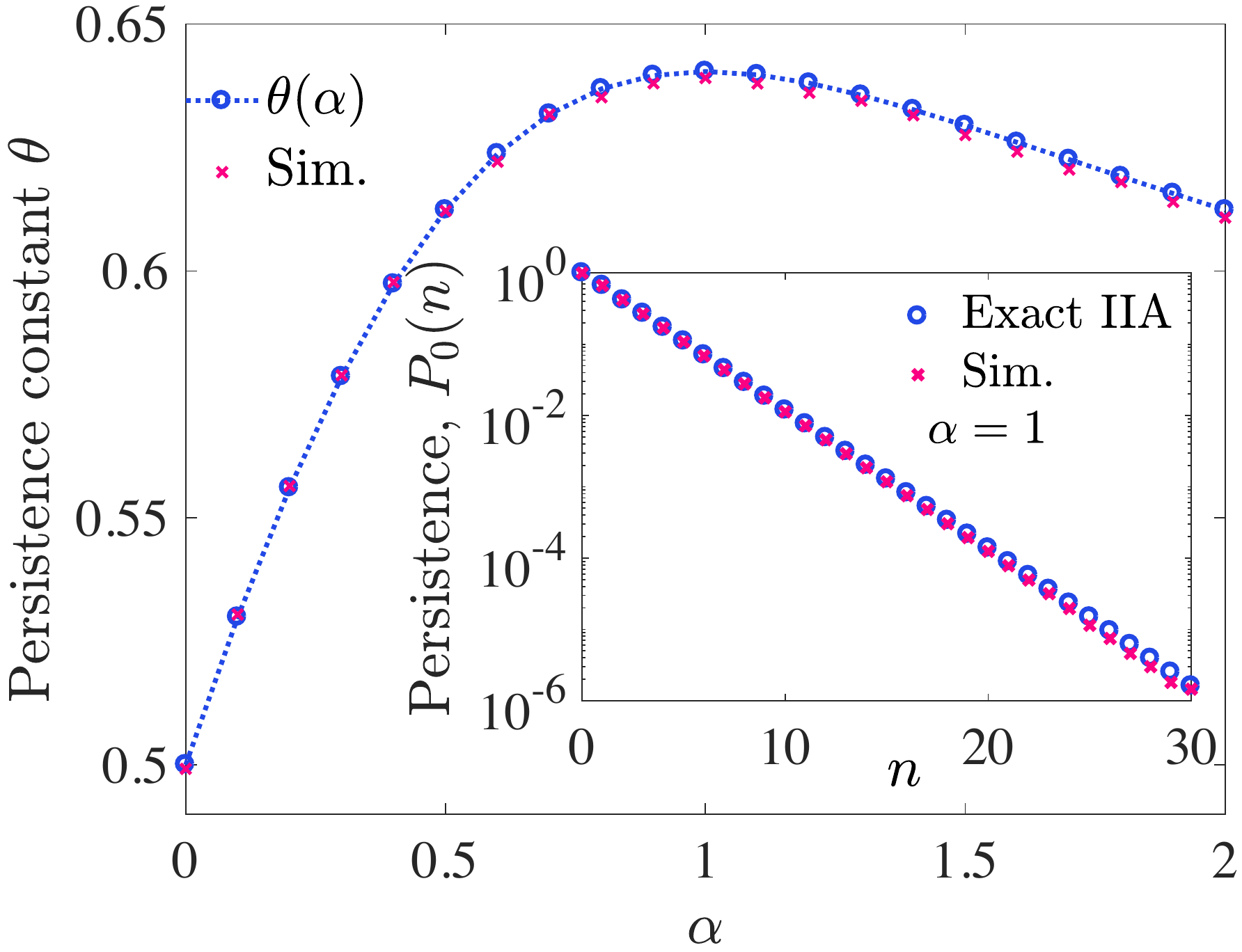}
	\caption{(color online). Persistence constant $\theta(\alpha)$ as a function of $\alpha$ for the process described by the equation of motion $x(n)=\eta(n)+\alpha\eta(n-1)$. Inset displays the persistence $P_0(n)$ for the case $\alpha=1$. Simulations are averaged over $10^6$ realizations. \label{fig:ma1}}
\end{figure}

\emph{Example 1}---The first GSP that we consider is the nearest neighbor-correlated \emph{first order moving average process} \cite{brockwell2006introduction}. It evolves for $n\geq1$ via 
\begin{equation}\label{eq:MA1}
x(n)=\eta(n)+\alpha\eta(n-1),
\end{equation}
where $\alpha$ is a constant and $\eta$ is Kronecker delta-correlated white noise $\langle\, \eta(n)\eta(k)\, \rangle=\sigma^2\delta_{n,k}$ with variance $\sigma^2$, which we set to unity. From Eq. \eqref{eq:MA1} the autocorrelator becomes
\begin{equation}\label{eq:MA1+corr}
A(n)=\begin{cases}
               1,\phantom{/(1-\alpha^2)}\quad\, n=0\\
               \alpha/(1+\alpha^2),\quad n=\pm1\\
               0,\phantom{/(1-\alpha^2)}\quad\, |n|>1
            \end{cases}
\end{equation}
and using Eq. \eqref{eq:MA1+corr} in Eq. \eqref{eq:om2+z} gives
\begin{equation}\label{eq:MA1+omega}
\omega(z)=\frac{r}{z}+\frac{1}{2z(z-1)}.
\end{equation}
This allows us to calculate the persistence and its constant exactly within the IIA. Using Eqs. \eqref{eq:persist+z} and \eqref{eq:MA1+omega}, we get
\begin{equation} \label{eq:MA1_invert1}
P_0(z)=\frac{z(1+4r(z-r))}{4r(z-z^*_+)(z-z^*_-)},
\end{equation}
with the poles
\begin{align} \label{eq:MA1_invert2}
z_{\pm}^*&=\frac{4 r-1\pm\sqrt{1 + 8 (1 - 2 r) r}}{8 r}.
\end{align}
We invert $P_0(z)$ with \cite{debnath2014integral}
\begin{equation} \label{eq:appc+2}
P_0(n)=\frac{1}{2\pi i}\oint_{\mathcal{C}}dz\,P_0(z)z^{n-1},
\end{equation}
where $\mathcal{C}$ is a positively oriented curve that encloses all poles. We have two simple poles at $z=z^*_{\pm}$, and  Cauchy's residue theorem therefore gives
\begin{equation} \label{eq:exact+ma1+pers}
P_0(n)=\frac{(z^*_+)^n(1+4r(z^*_+-r))-(z^*_-)^n(1+4r(z^*_--r))}{4r(z^*_+-z^*_-)}.
\end{equation}
From this we identify the slowest decaying term that depends on $n$ as the persistence constant $\theta(\alpha)=z^*_+$. Thus
\begin{equation} \label{eq:theta+alpha}
\theta(\alpha)=\frac{4 r-1+\sqrt{1+8 (1-2 r) r}}{8r} ,
\end{equation}
where $\alpha$ enters through $r=\cos^{-1}\left( \alpha/(1+\alpha^2) \right)/\pi$. For $\alpha=0$, the process is Markovian and the probability of making a sign change at each step is $1/2$, yielding the trivial asymptotic behavior, $P_0(n)\sim2^{-n}$. When $\alpha=0$, the IIA becomes exact as each interval between sign changes is uncorrelated. 
Indeed, $\alpha=0$ ($r=1/2$) in Eq. \eqref{eq:theta+alpha} gives $\theta=1/2$.

Equation \eqref{eq:theta+alpha} is new, but approximative for $\alpha\neq0$. With $\alpha=1$, ($r=1/3$) in Eq. \eqref{eq:theta+alpha} we get $\theta(1)=(1+\sqrt{17})/8\approx 0.6404$, which is close to the exact result $2/\pi\approx 0.6367$ \cite{majumdar2001persistence2}. For other values of $\alpha$, we compare Eq. \eqref{eq:theta+alpha} to simulations (see Fig. \ref{fig:ma1}) where the inset displays the result for $P_0(n)$ in Eq. \eqref{eq:exact+ma1+pers} for $\alpha=1$. In all aspects, we see good results compared to simulations.
\begin{figure}[] 
	\includegraphics[width=1\columnwidth]{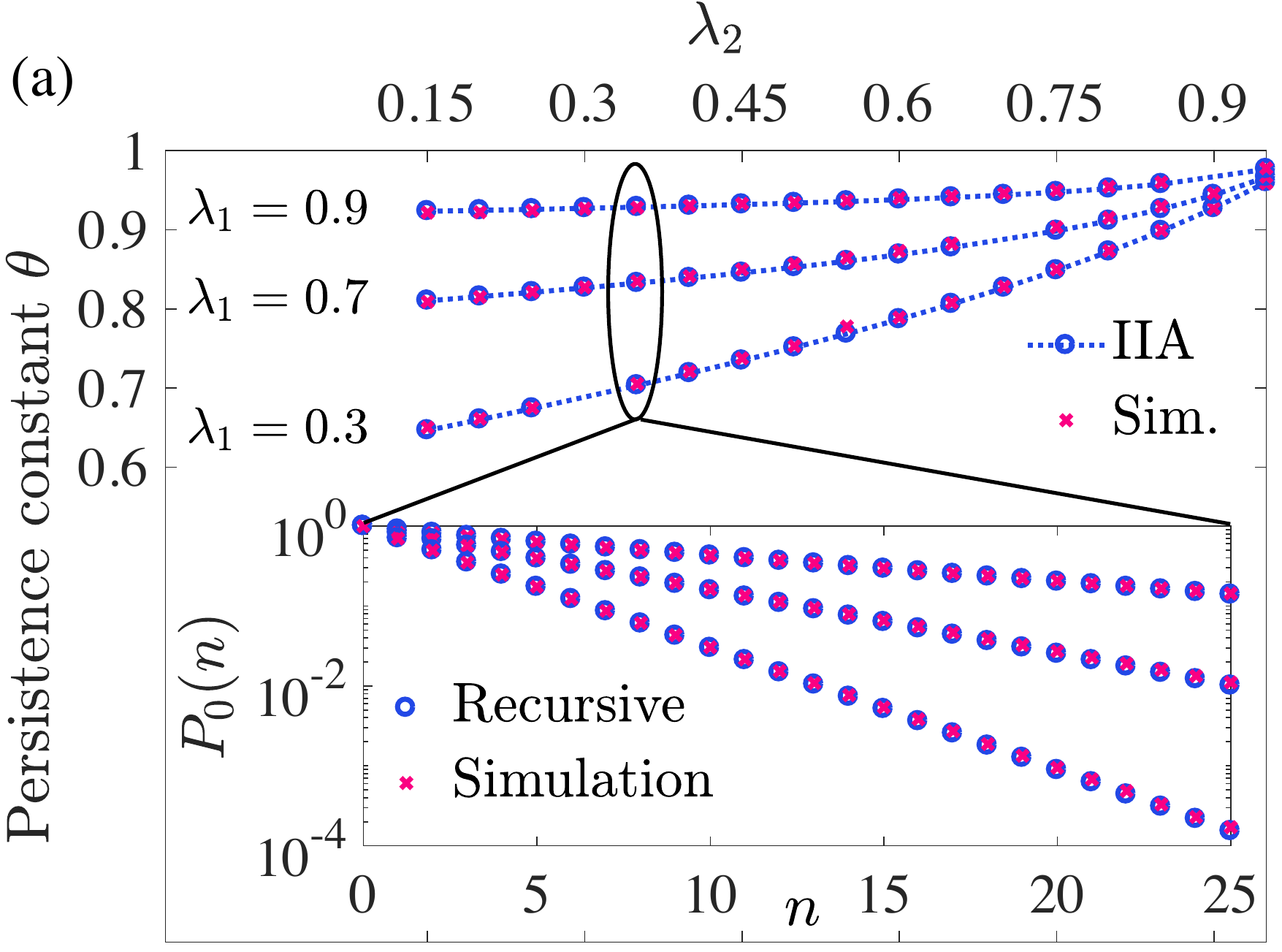}
	\includegraphics[width=1\columnwidth]{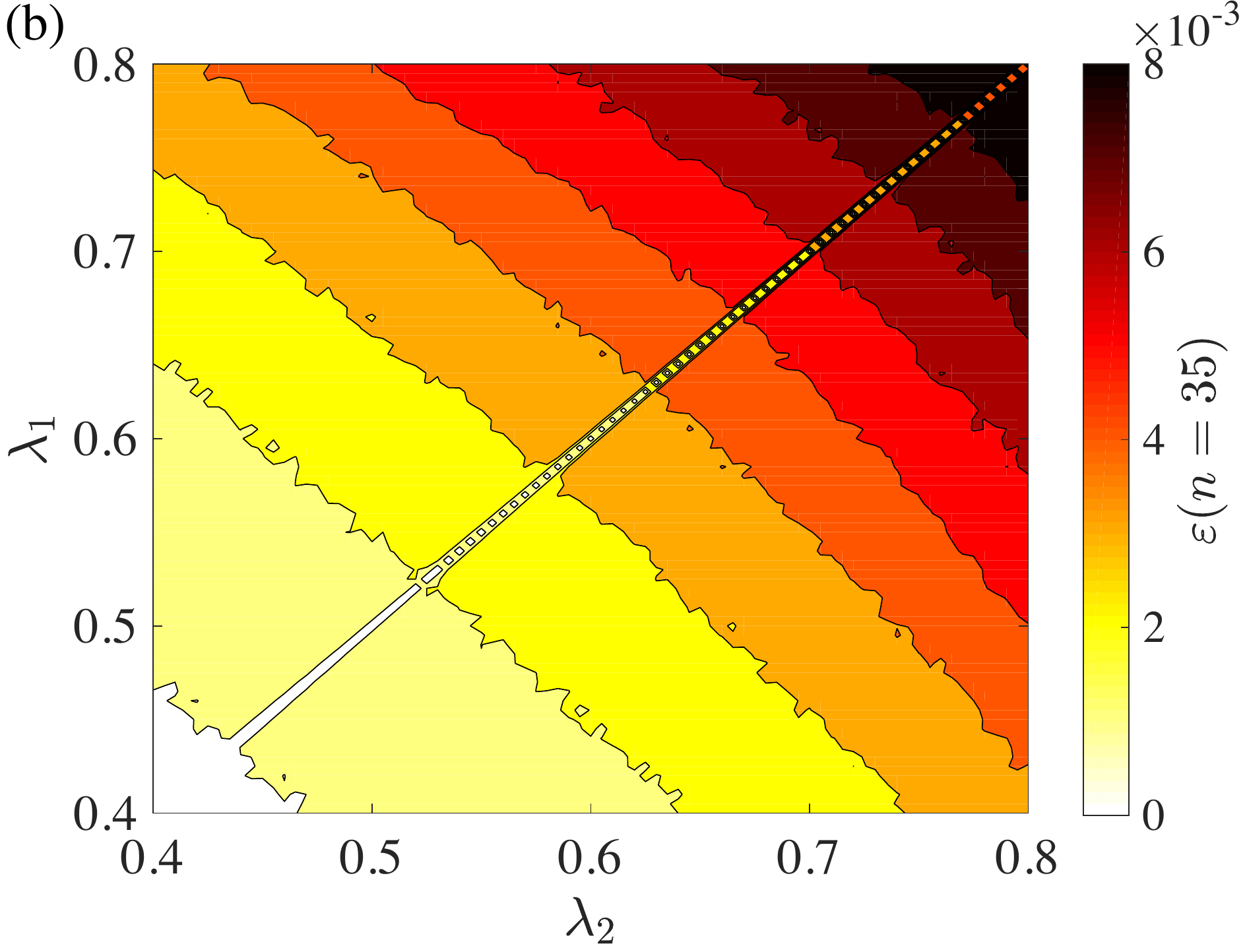}
	\caption{(color online). (a): Persistence constant $\theta$ for the two-step memory process, $x(n)=\phi_1x(n-1)+\phi_2x(n-2)+\eta(n)$, for different values of $\lambda_{1,2}$ ($\phi_{1,2}$). Inset displays the persistence probability $P_0(n)$ for fixed $\lambda_2=0.35$. Simulations are averaged over $10^6$ realisations. (b): Heat map displaying the mean absolute error of the persistence up to $n=35$: recursive formula vs. simulation. Simulation is averaged over $4\times10^6$ realisations. In both (a) and (b), note that the points where $\lambda_1=\lambda_2$ are excluded, as they make the autocorrelator divergent. \label{fig:AR2_pers}}
\end{figure}

\emph{Example 2}---In Example 1 we saw that the IIA can be successfully applied to nearest neighbor-correlated variables. To increase the process' complexity, we next consider variables that have an exponential decaying correlation. The simplest non-Markovian member of this class is the \emph{second order autoregressive process} \cite{brockwell2006introduction}. It is a two-step memory process governed by the equation of motion
\begin{equation}\label{eq:AR2}
x(n)=\phi_1 x(n-1)+\phi_2 x(n-2)+\eta(n),
\end{equation}
for $n\geq2$ with $\phi_{1,2}$ constants. The autocorrelator is given by (see Appendix \ref{app:b} for details)
\begin{equation}\label{eq:AR2+corr}
A(n)=\frac{\lambda_1^{n+1}(1-\lambda_2^2)-\lambda_2^{n+1}(1-\lambda_1^2)}{(\lambda_1-\lambda_2)(1+\lambda_1\lambda_2)},
\end{equation}
where $\lambda_{1,2}$ depends implicitly on $\phi_{1,2}$ through the relations $\phi_1=\lambda_1+\lambda_2$ and $\phi_2=-\lambda_1\lambda_2$ with $\lambda_1\neq\lambda_2$ and $|\lambda_{1}|,|\lambda_{2}|<1$, which puts boundaries on the values of $\phi_{1,2}$. 

Using Eqs. \eqref{eq:roots+theta} and \eqref{eq:om2+z}, we numerically solve for the persistence constant $\theta$. In Fig. \ref{fig:AR2_pers} (a), we plot $\theta$ vs. $\lambda_2$ for three different values of $\lambda_1$. In the inset, we show the results from the recursive relation in Eq. \eqref{eq:persist+n} with fixed $\lambda_2=0.35$. 

Using the recursive formula in Eq. \eqref{eq:persist+n}, we also show in Fig. \ref{fig:AR2_pers} (b) the mean absolute error, $\varepsilon(n)=\sum_{k=0}^n|P_0^{\text{Sim}}(k)-P_0^{\text{Rec}}(k)|/(n+1)$, with the expected behavior: as the autocorrelator decays faster $(\lambda_{1,2}\to0)$, $\varepsilon$ decreases, and vice versa. Note that $A(n)$ and $\phi_{1,2}$ are invariant when interchanging $\lambda_1\rightleftarrows\lambda_2$, which is why the heat map is symmetric along the diagonal. In all aspects, we see good results compared to simulations.  

\begin{figure}[] 
	\centering
	\includegraphics[width=1\columnwidth]{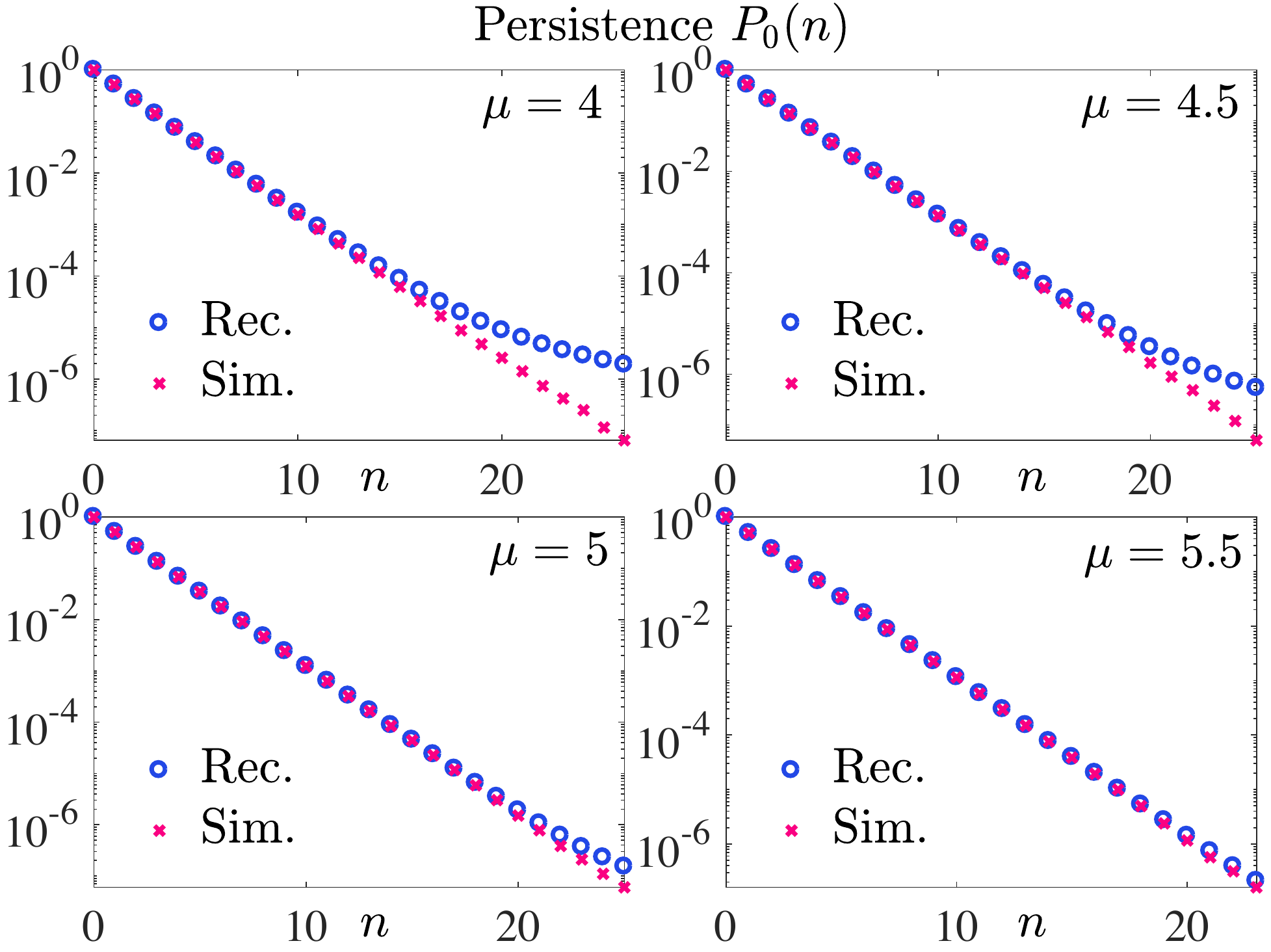}
	\caption{(color online). Persistence probability $P_0(n)$ (on the ordinate), simulation vs. recursive formula. Simulations plotted as far as they have converged, averaged over $10^8$ realisations.  \label{fig:power}}
\end{figure}

\emph{Example 3}---We showed in the above examples that the IIA is a good method when dealing with weakly correlated variables. However, where is the limit where the variables become too strongly correlated and the IIA breaks down? To investigate this, we consider an extreme case with a power law autocorrelator given by
\begin{equation} \label{eq:power_law}
\begin{aligned}
A(n)&=\left(  1+n \right)^{-\mu}\sim n^{-\mu}, \\
\end{aligned}
\end{equation}
and we want to find the smallest $\mu$ where the simulated persistence and the recursive formula in Eq. \eqref{eq:persist+n} agree down to $n=20$. The result is displayed in Fig. \ref{fig:power}. By inspection, we see that, for $\mu\gtrsim5$, we match the simulations well down to $n=20$. For $\mu<5$ we start to see deviations between the recursive formula and the simulations at $n<20$. However, these deviations occur much later than the mean time $\langle n \rangle$, where the first sign change occur. For example, when $\mu=4$ then $\langle n \rangle\approx 2.1$ (see Tab. \ref{tab:mfpt}), which means that most trajectories will have changed their sign long before deviations are substantial.  This result can be compared to fractional Gaussian noise \cite{ibe2013elements}, which exhibits the power law decay $A(n)\sim n^{2H-2}$ for large $n$, $H$ being the Hurst index, $0<H<1$. Thus, translated to the exponent $\mu$, we have values between $0<\mu<2$ for our toy model. Therefore, we conclude that strongly correlated variables, like fractional Gaussian noise, is not applicable to our results. However, to generalize the IIA to these kinds of processes is a big challenge that goes beyond the scope of this work.

\textbf{\emph{Mean first-passage time}.---}As a sub-result, we calculate the mean first-passage time till the first sign change, $\langle n \rangle$. Naively, and guided by Rice's formula, one might guess that $\langle n \rangle\approx1/r$, as this gives a measure of the time needed before the process changes sign. However, this does not take the memory of the process into consideration, since the probability that $x(n)$ changes sign will in general depend on all the steps leading up to time $n$. To calculate $\langle n \rangle$ within the IIA, it is useful to define the auxiliary function 
\begin{equation} \label{eq:om2+z+y}
y(z)=\frac{2}{z\pi}\sum_{n=0}^{\infty}\sin^{-1}(A(n))z^{-n},
\end{equation}
which, together with Eqs. \eqref{eq:fptd+z} and \eqref{eq:om2+z}, gives a compact expression for the first-passage time density
\begin{equation}\label{eq:fptd+z2}
\rho(z)= \frac{1-(z-1)y(z)}{1+\frac{z-1}{2r}\left[1-(z-1)y(z)    \right]}.
\end{equation}
The mean first-passage time can be calculated from $\langle n \rangle=\sum_{n=0}^{\infty}n\rho(n)=-d\rho(z)/dz\big|_{z=1}$, which leads to the summation formula
\begin{equation}\label{eq:fptd+mfpt}
\langle n \rangle = \frac{1}{2r}+\frac{2}{\pi}\sum_{k=0}^{\infty}\sin^{-1}(A(k)),
\end{equation}
that is different from the naive assumption $1/r$. We get good agreement compared to simulations for the processes discussed herein, see numerical values in Tab. \ref{tab:mfpt}.
\begin{table}[]
\begin{center}
\caption{Comparing Eq. \eqref{eq:fptd+mfpt} and simulations (averaged over $10^7$ realisations) for the mean first-passage time. The autocorrelator $A(n)$ used is indicated by its parameters in the left-most column. \label{tab:mfpt}}
  \begin{tabular}{ | l || c | c |}
    \hline    
    $A(n)$ & $\langle n \rangle$ (Eq. \eqref{eq:fptd+mfpt}) & $\langle n \rangle$ (Sim.) \\ \hline \hline
    $\alpha = 1$ & 2.8333 & 2.8172 \\ \hline
    $\lambda_1=0.9, \lambda_2=0$ & 10.6221 & 10.5276 \\ \hline
    $\lambda_1=0.9, \lambda_2=0.5$ & 14.1929 & 14.5686 \\ \hline
    $\mu = 4.5$ & 2.0638 & 2.0627 \\ \hline
    $\mu = 4$ & 2.0939 & 2.0910 \\
    \hline
  \end{tabular}
\end{center}
\end{table}

\textbf{\emph{Summary \& Discussion}.---}For a Gaussian stationary process (GSP) in discrete time $n$, the persistence probability for large times $n$ is characterized by the persistence constant $\theta$ through $P_0(n)\sim\theta^n$. In general, $\theta$ is non-trivial to calculate except for a few special cases such as $x(n)=\eta(n)$ (see Eq. \eqref{eq:MA1} with $\alpha=0$). To tackle this problem, we provide a simple method based on the independent interval approximation (IIA), where we have derived a new set of equations for calculating the persistence constant but also the full persistence probability for any time $n$ via a recursive formula that is valid for a general GSP.

When analyzing data from measurements and simulations, it is important to respect that the data is a collection of discretely sampled numbers. Thus, using a persistence theory based on continuous time, effectively approximating the discrete time process with a continuous one, the persistence probability might be overestimated. This is because the continuous process may change sign an even number of times between two consecutive discrete time points, which will not happen in a discrete theory.

Nevertheless, most results for the persistence probability are in continuous time $t$. For a general non-Markovian GSP with autocorrelator $A(t)$, there are no exact results except if $A(t)<1/t$ for large $t$, then $P_0(t)\sim \text{exp}\left(-\theta_c t  \right)$ \cite{majumdar1999persistence}. To find $\theta_c$, there are several approximations where the IIA is one of the most successful methods that can be applied to a wide range of smooth processes \cite{bray2013persistence}. Our work can be seen as an extension to that. Indeed, with the time increment $\Delta t$,  we set $t_n=n\Delta t$,  and keep $n\Delta t$ fixed as we let $\Delta t\to0$ and $n\to\infty$. This is the continuum limit of our equations. In this limit, we replace the sums in Eq. \eqref{eq:pmn+iia} with integrals, and the Rice rate $r$ is replaced by $r_c=\sqrt{-A''(t_0)}/\pi$ \cite{rice1944mathematical}. Then one proceeds as in discrete time, but using the Laplace transform, $\mathscr{L}\left\{f(t)\right\}=f(s)$, instead of the $z$-transform. It is then possible to find the persistence constant $\theta_c$  numerically as the first root on the negative $s$-axis from the equation $1+\frac{s}{2r_c}\left(1-\frac{2s}{\pi}\mathscr{L}\left\{\sin^{-1}(A(t))\right\}  \right)=0$, since $\mathscr{L}\left\{\text{exp}\left(-\theta_c t  \right)\right\}=1/(s+\theta_c)$. This is the continuous time version of Eq. \eqref{eq:roots+theta}, and it is also found in e.g. \cite{majumdar1999persistence}.

Prior to our work, the main method for calculating  the persistence constant $\theta$ for a general GSP in discrete time has been via a series expansion in terms of the autocorrelator \cite{ehrhardt2002series}. To the 14th order, automated with a computer, the authors in \cite{ehrhardt2002series} found good numerical results for $\theta$ for weakly correlated variables. While the IIA also relies on weakly correlated variables it can not be systematically improved, compared to, e.g., a series expansion. However, our work is less involved and provides closed-form expressions, arguably simpler expressions than \cite{ehrhardt2002series}, and a recursive formula for the full persistence probability. Comparing the persistence constant for the non-Markovian process $x(n)=\eta(n)+\eta(n-1)$ to the exact result, our IIA approach is identical down to two significant figures.

The persistence $P_0(n)$ is just the special case $m=0$ of the more general probability distribution $P_m(n)$ that $m$ sign changes have occurred up to $n$. The only exact result related to $P_m(n)$ is its first moment $\langle m(n) \rangle$, given by Rice's result [see Eq. \eqref{eq:rice}]. While the full distribution still is unknown, for large $n$ it tends to a Gaussian, $P_m(n)\sim\text{exp}\left[ -(m-\langle m(n) \rangle)^2/2\sigma^2(n) \right]$ \cite{ho1987central}, characterized by its first two cumulants, $\langle m(n) \rangle$ and $\sigma^2(n)$, which are process-specific. Advancements have been made in this regime of the already mentioned process $x(n)=\eta(n)+\eta(n-1)$, using large deviation theory \cite{majumdar2002statistics} but also in \cite{ehrhardt2004persistence}, where $\sigma^2(n)$ was calculated using the same expansion technique as in \cite{ehrhardt2002series} for weakly correlated variables. More recently, higher order cumulants (and moments) of $P_m(n)$ were calculated using the IIA with good results for higher order autoregressive processes \cite{nyberg2017}.
In the future, it would be interesting to see to what extent the IIA can be used to find the full $P_m(n)$.

\textbf{\emph{Acknowledgement}.---}MN and LL wish to thank Tobias Ambj\"{o}rnsson for inspiring discussions and acknowledge financial support from the Swedish Research Council (Vetenskapsr\aa det, Grant No. 2012-4526).


\newpage
\appendix
\section{Formal inversion of the first-passage time density} \label{app:a}
In the main text, to invert Eq. \eqref{eq:fptd+z} and retrieve Eq. \eqref{eq:fptd+n}, we use that
\begin{align*}
Z\left\{ f(n+k) \right\}&=z^kf(z)-z^kf(0)\\
&-z^{k-1}f(1)-\ldots -zf(k-1)
\end{align*}
and
\begin{align*}
\omega(1)=r.
\end{align*}
Together with the forward difference operator, $\Delta f(n)=f(n+1)-f(n)$, we rewrite Eq. \eqref{eq:fptd+z} as
\begin{align} \label{eq:app+fptd+z}
\rho(z)Z\left\{ \Delta^2\omega(n) \right\}+2rz\rho(z)=2rZ\left\{ \Delta\omega(n) \right\},
\end{align}
where $Z\left\{ f(n) \right\}=f(z)$. With $Z^{-1}Z\left\{ f(n) \right\}=f(n)$ and the convolution property $Z\left\{ \sum_{j=0}^n f(j)g(n-j) \right\}=f(z)g(z)$, we apply $Z^{-1}$ from the left to Eq. \eqref{eq:app+fptd+z} which yields
\begin{align} \label{eq:app+fptd+z2}
\sum_{j=0}^n\rho(j)\Delta^2\omega(n-j)+2r\rho(n+1)=2r\Delta\omega(n),
\end{align}
where we used the initial conditions $\rho(0)=\omega(0)=0$. Finally, re-arrangement of terms gives Eq. \eqref{eq:fptd+n} in the main text.

\section{Probability of an odd number of sign changes} \label{app:c}
To calculate the probability that an odd number of sign changes has occurred, $\omega(n)$, we start from the conditional probability density function of the Gaussian stationary process (GSP), given by \cite{brockwell2006introduction}
\begin{equation} \label{eq:cond+pdf}
\begin{aligned}
P(x_n|x_0) &= \frac{\text{exp}\left( -\frac{(x_n-A(n)x_0)^2}{2\gamma(0)(1-A(n)^2)} \right)}{\sqrt{2\pi \gamma(0)(1-A(n)^2)}} ,
\end{aligned}
\end{equation}
where $\gamma(n)=\langle\,x(n+k)x(k)\,\rangle$ is the covariance and $A(n)=\gamma(n)/\gamma(0)$ is the autocorrelator. Using Eq. \eqref{eq:cond+pdf}, we can calculate the probabilities that $x(n)$ is above and below the zero given the initial position $x_0$. They are given by
\begin{equation} \label{eq:om1}
\begin{aligned}
\omega^+(n|x_0)&=\int_{0}^{\infty} dx_nP (x_n|x_0)\\ &= \frac{1}{2}\text{erfc}\left(\frac{-x_0A(n)}{\sqrt{2\gamma(0)[1-A(n)^2]}}\right),
\end{aligned}
\end{equation}
and
\begin{equation} \label{eq:om2}
\begin{aligned}
\omega^-(n|x_0)&=\int_{-\infty}^0 dx_nP (x_n|x_0)\\ &= \frac{1}{2}\text{erfc}\left(\frac{x_0A(n)}{\sqrt{2\gamma(0)[1-A(n)^2]}}\right),
\end{aligned}
\end{equation}
respectively. Since we work in the stationary limit, we want to average these quantities over the equilibrium density $g(x)$, found from Eq. \eqref{eq:cond+pdf} as $n\to\infty$,
\begin{equation} \label{eq:equi+dens}
g(x)=\frac{1}{\sqrt{2\pi\gamma(0)}}\text{exp}\left( -\frac{x^2}{2\gamma(0)}  \right).
\end{equation}
Thus, the probability of having an odd number of sign changes at time $n$, in the stationary limit, is then given by
\begin{equation} \label{eq:om}
\begin{aligned}
\omega(n)&=\int_{0}^{\infty} dx_0\omega^-(n|x_0)g(x_0)+ \int_{-\infty}^{0} dx_0\omega^+(n|x_0)g(x_0) \\ &=\frac{1}{2}-\frac{1}{\pi}\sin^{-1}(A(n)),
\end{aligned}
\end{equation}
which is Eq. \eqref{eq:omega_n} in the main text. The solution to this integral is also found in \cite{owen1980table}.

\section{Autocorrelator for the autoregressive process of order two} \label{app:b}
The autocorrelator of the process in Eq. \eqref{eq:AR2} in the main text can be found by defining the backward operator $B$ as: $B^jx(n)=x(n-j)$. If $\phi(B)=1-\phi_1 B-\phi_2 B^2$, then Eq. \eqref{eq:AR2} can be written $x(n)=\eta(n)/\phi(B)$. Next we set $\phi(B)=(1-\lambda_1B)(1-\lambda_2B)$ and identify that $\phi_1=\lambda_1+\lambda_2$ and $\phi_2=-\lambda_1\lambda_2$. Using partial fraction gives
\begin{align} \label{eq:app+ar2+1}
x(n)=\left( \frac{c_1}{1-\lambda_1B}+\frac{c_2}{1-\lambda_2B}  \right) \eta(n),
\end{align}
where $c_1=\lambda_1/(\lambda_1-\lambda_2)$ and $c_2=\lambda_2/(\lambda_2-\lambda_1)$ with $\lambda_1\neq\lambda_2$. For $|\lambda_{1,2}|<1$, we expand Eq. \eqref{eq:app+ar2+1}
\begin{equation} \label{eq:app+ar2+2}
\begin{aligned}
x(n)&=\sum_{j=0}^{\infty}\left(c_1\lambda_1^jB^j +c_2\lambda_2^jB^j \right) \eta(n) \\
&=\sum_{j=0}^{\infty}\left(c_1\lambda_1^j +c_2\lambda_2^j \right) \eta(n-j).
\end{aligned}
\end{equation}
Taking the expectation value $\gamma(n)=\langle\, x(n+k)x(k) \,\rangle$ and assuming that the noise $\eta(n)$ is Kronecker delta-correlated, $\langle \,\eta(n_1)\eta(n_2) \,\rangle=\sigma^2\delta_{n_1,n_2}$, gives
\begin{equation} \label{eq:app+ar2+3}
\gamma(n)=\sigma^2\frac{\frac{\lambda_1^{n+1}}{1-\lambda_1^2}-\frac{\lambda_2^{n+1}}{1-\lambda_2^2} }{(\lambda_1-\lambda_2)(1-\lambda_1\lambda_2)},
\end{equation}
which yields the autocorrelator $A(n)$ in Eq. \eqref{eq:AR2+corr} in the main text via $A(n)=\gamma(n)/\gamma(0)$.

\end{document}